# Medusa: Blockchain Powered Log Storage System

Hao Wang
*Humsen*
Beijing, *China*
haow85@live.com

Desheng Yang, Nian Duan, Yang Guo and Lu Zhang
*Humsen*
Beijing, *China*
{yangdesheng, duannian, guoyang, zhanglu}@hansheng.io

*Abstract*—**Blockchain is one of the most heavily invested technologies in recent years. Due to its tamper-proof and decentralization properties, blockchain has become an ideal utility for data storage that is applicable in many real world industrial scenarios. One important scenario is web log, which is treated as sources of technical significance and commercial revenues in major internet companies. In this paper, we illustrate our design of a web log storage system based on HyperLedger. HyperLedger yields higher throughput and lower latency compared with other blockchain systems. Alongside its efficiency advantages, HyperLeger is a permissioned blockchain, which is an ideal fit for enterprise software design scenario.**

## I. INTRODUCTION

Blockchain is an emerging technology that is rapidly changing the landscape of many industries. Due to its ease-of-use, tamper-proof, and decentralization properties, it is a suitable candidate for many industrial engineering scenarios. Major financial institutions have continued investing heavily in blockchain development. Internet companies are also in hurry to take advantage of blockchains to lead the next wave of innovation. Blockchain is the hottest venture capitalist investment sector in 2018 in China.

Big data has drawn a lot of traction in last ten years. Web 2.0 empowered people's ability to produce and share data. Consumer companies and internet firms have exploited all kinds of technologies related to data science to generate customer insight or facilitate AI processes to yield revenues. Mature business models combining big data and AI are commonplaces across the globe. However, how to combine the power of blockchain and big data has mostly been ideas incubated in research labs and educational institutions. There is still a long way to go before commercial blockchain plus big data solutions become available.

Logs are important data assets within companies and government agencies. Effective and secure utilization of logs is in compliance with the concept of data governance which advocates the proper use of data throughout its life cycle. However, due to high latency and low throughput, blockchains could not withhold tens of thousands of TPS commonly seen with Nginx servers. It is unlikely to get blockchains to replace a mature real-time log storage and analytics engine like ElasticSearch in the near future. In spite of its technical limitation, blockchain could serve as a storage option for audit logs in consumer companies and government agencies. Such logs are generated with low frequency but need high level security and they have to be tamper-proof.

In this paper, we illustrate an audit log storage system we developed with the name Medusa. We created our system with one of the most tested and popular blockchain framework HyperLedger Fabric. Our audit log storage system could serve as a precursor to more commercially ready solutions for wider audience in the industry.

In the current stage of development , our system could not withhold high TPS due to intrinsic technical limitations of blockchains, but Medusa yields good performance in low-latency and non-realtime batch processing scenarios. Medusa takes advantage of the smart contract functionality of HyperLedger. It is easy to deploy and use for end users.

## II. RELATED WORK

Blockchain and Bitcoin [1] were invented as twins in 2009. Since its invention, blockchain and cryptocurrencies have released torrents of energy that has revolutionized the financial industry. However, blockchain technology is still in its embryo. The number of research papers and patents are highly limited compared with other computer science fields. One major bottleneck for the wide adoption of blockchain technology is its low latency compared with conventional centralized IT systems [2]. For example, the blockchain underlying Bitcoin generates a block every 10 minutes. Such speed is too slow for many real world business scenarios.

One factor that affects the speed of blockchain is its consensus protocol. Consensus protocols have been researched for decades by researchers in other fields such as distributed computing systems. For a long time in the past, the majority of Byzantine algorithms had no sub-exponential computational complexity [2][3][4] without worsening other conditions such as the number of tolerable malicious nodes [5] . Effective consensus protocols is a popularly researched topic today with many new algorithmic inventions [6][7].

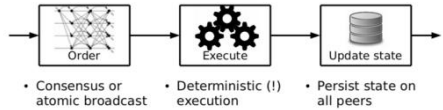

Fig 1. Order-execute architecture of conventional blockchains [11]

Log auditing is a critical process in commercial companies and government agencies. There are multiple categories of logs. Web logs generated by Nginx or Tomcat servers are one of the major research targets of data scientists. Important business systems such as recommender systems [8], fraud detection systems [9][10] and computational advertisement systems take web logs as their input data sources. However, due to the high throughput and low latency of web servers , blockchain systems could not withhold the voluminous data transactions generated in the short period of time. Despite its infeasibility as commercial technical solutions,  blockchain technology is an ideal choice for log auditing due to its tamper-proof property. Unlike commercial technical solutions, log auditing has a lower requirement for computational efficiency, therefore eliminating the technical limitation imposed by blockchains' low speed. [11]

III. PERMISSIONED BLOCKCHAIN

The first blockchain system supporting Bitcoin is a permission-less blockchain. A permission-less blockchain is a chain where all users remain anonymous on the chain network. In such a blockchain paradigm, transaction validation process preventing nefarious behaviors is costly due to users' undisclosed identities. For example, it takes Bitcoin 10 minutes to generate a block. Such computational speed is impractical for most real world applications where conventional solutions could support tens of thousands of QPS and milliseconds latency.

A permissioned blockchain system is a blockchain system where users have identities on the network. In permission-less blockchain systems, a schema called cryptocurrency mining is invented to provide incentives for the transaction validation work. However, in a permissioned blockchain system, the cryptocurrency mining process is eliminated and the computational speed of the system is a lot faster than permission-less blockchains.

Since permission-less blockchains such as Bitcoin suffer from bad user experiences due to its computational speed limitation. We chose a permissioned blockchain network HyperLedger for our log storage system design.  In the following section, we illustrate the basic concepts and architectures behind HyperLedger.

IV. HYPERLEDGER

The majority of blockchains including the popular ETH blockchain developed prior to HyperLedger deploys an order-execute schema which makes their speed intolerably slow for commercial applications. In order-execute schema, transactions are ordered sequentially on every peer of the network before their final execution. Every peer in the network has to wait for the arrival of all transactions. This process could not be parallelized, which is the bottleneck of the entire computational process.

Contrary to the order-execute schema, HyperLedger adopted an execute-order-validate schema . Each stage of schema could be processed separately, leaving room for computational distribution. Each transaction in the system is endorsed first, validated with conflict check second and executed in the end.

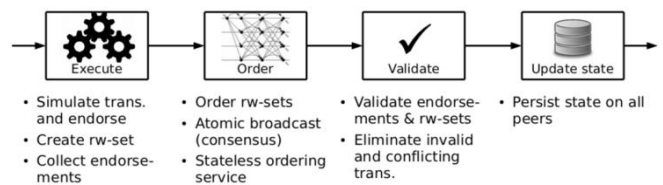

Fig 2. Execute-order-validate architecture of HyperLedger [12]

HyperLedger is a blockchain network where each blockchain runs in a separate docker container and segregated from other blockchains. A blockchain in HyperLedger is called a channel, where participants in one channel could not interact with participants in other channels. HyperLedger has a very well supported smart contract functionality, making it suitable in many financial application scenarios. The high level transaction flow chart of HyperLedger is shown in Fig. 3

V. LOG STORAGE

Conventional log storage systems include Infobright , Greenplum and other data warehousing solutions. As the internet entered the big data era, HDFS and other distributed file systems become common options for log file storage, which are prevalent in commercial companies and government agencies.

A log storage system for critical data stores data permanently and is tamper-proof. Such log storage requires two major operations: data appending and data querying. Data modification is infeasible due to the intrinsic properties of blockchains. Since Medusa supports only critical data storage rather than holistic big data support, simplistic operations are enough in most user cases. Therefore we also support only 2 operations : data appending and data querying.

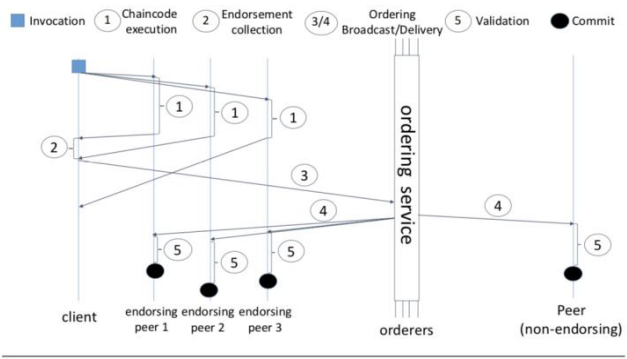

Fig 3. High-level transaction flow of HyperLedger Fabric

## VI. SYSTEM DESIGN

Our system is based on HyperLedger. In HyperLedger, there are 3 elements in a blockchain network, namely participants, assets and transactions. These building elements are foundations of HyperLedger smart contract functionality. HyperLedger also has built-in functionality for queries.

In our system, there is only one type of participants called DataSource. A DataSource is the owner of data. It could be the source or the sink of audit logs. DataSource is indexed by datasourceId, it has 5 attributes: ip, port, username, password and url. Ip, port and url specifies how to access the source or sink of the audit logs, where username and password serve as authentication measures.

There is also only one type of asset, which is WebLogData. In our system, we took web log data as an example to specify the asset's properties, even though in reality there are many different data formats. The asset's properties other than the ID are url, referer , returnCode, userAgent, datetime and IP. WebLogData could be treated as a simplified simulation of real web log data.

Since our log system is tamper-proof, the sole operation allowed in the system is log appending. We define a transaction type called DataAppend that deals with the operation. DataAppend has one attribute, namely the data to append to the blockchain. We define a trigger function onDataAppend that writes the appending data to the blockchain when the event is triggered.

For the query functionality of our system, we define the selectWebLogData function that retrieves all the log data with a list of retrieval functions that fetch data based on log data attributes such as IP, user agent and date time.

HyperLedger has enabled the simplicity of our system as just mentioned above. We developed our HyperLedger code with HyperLedger Composer and deployed our system on HyperLedger Fabric Network.

## VII. CONCLUSION

Blockchain is one of the most cutting-edge technologies in recent years. Its tamper-proof and decentralization properties allow enterprises especially financial institutions to execute business processes with privacy and security. Researchers have hypothesized on how to combine the power of blockchain and big data together. However, the hypothesis has not been successfully materialized due to intrinsic technical limitation of blockchains. In this paper, we demonstrated our log storage system based on HyperLedger. Our system is designed for log data auditing on critical data of enterprises and government agencies.

Most blockchains are notoriously slow for practical applications. Both Bitcoin and Ethereum systems are not suitable competitors when compared with conventional centralized IT systems. HyperLedger as a distributed ledger has boosted the TPS of transactions to thousands from the single digit number of Bitcoin. In addition to its high efficiency and scalability, HyperLedger provides user friendly programming interfaces that is easy to learn and use. These are the major factors behind our choice of HyperLedger smart contracts as our underlying technical paradigm for Medusa.

Although HyperLedger is more scalable and efficient than many other blockchains, it is still far behind the capacity of conventional IT solutions such as ElasticSearch in the field of data storage systems. Building blockchain powered storage system that's fast and scalable with security is our long term goal.

In future work, we wish we could design a full-feature log system that tolerates higher throughput with lower latency. Our final goal is to get as close as possible to a blockchain powered log system similar to ElasticSearch that works for not only auditing logs but also web server logs. Our work in this paper serves as a precursor to our hypothesized system in the future.